\pdfoutput=1
\documentclass[journal]{IEEEtran}
\usepackage{amsmath}
\DeclareMathOperator*{\argmax}{arg\!\max}
\DeclareMathOperator*{\argmin}{arg\!\min}
\usepackage{float}
\usepackage{graphicx}
\usepackage{epstopdf}
\makeatletter
\def\footnoterule{\kern-3\p@
  \hrule \@width 2in \kern 2.6\p@} 
\makeatother
\ifCLASSINFOpdf
\else
\fi
\begin{document}

%
\title{ MimickNet, Matching Clinical Post-Processing Under Realistic Black-Box Constraints }

\author{Ouwen~Huang,~\IEEEmembership{Student Member,~IEEE},
        Will~Long,~\IEEEmembership{Student Member,~IEEE},
        Nick~Bottenus,
        Gregg~E.~Trahey,~\IEEEmembership{Senior Member,~IEEE},
        Sina~Farsiu,~\IEEEmembership{Senior Member,~IEEE},
        and~Mark~L.~Palmeri,~\IEEEmembership{Senior Member,~IEEE}
\thanks{
This work has been submitted to the IEEE Transactions on Medical Imaging on July 1st, 2019 for possible publication. Copyright may be transferred without notice, after which this version may no longer be accessible. It was supported by the National Institute of Biomedical Imaging and Bioengineering under Grant R01-EB026574 and National Institute of Health under Grant 5T32GM007171-44.

O. Huang, W. Long, N. Bottenus, G. E. Trahey, S. Farsiu, and M. Palmeri are with the Department of Biomedical Engineering at Duke University, Durham, NC 27708 USA (e-mail: ouwen.huang@duke.edu; willie.long@duke.edu; nick.bottenus@duke.edu; gregg.trahey@duke.edu; sina.farsiu@duke.edu; mark.palmeri@duke.edu). 
}
}

%
%

\markboth{Preprint}%
{Ouwen \MakeLowercase{\textit{et al.}}: MimickNet, Mimicking Clinical Image Post-Processing under black-box Constraints}
%



\maketitle

\begin{abstract}
Image post-processing is used in clinical-grade ultrasound scanners to improve image quality (e.g., reduce speckle noise and enhance contrast). These post-processing techniques vary across manufacturers and are generally kept proprietary, which presents a challenge for researchers looking to match current clinical-grade workflows. We introduce a deep learning framework, MimickNet, that transforms raw conventional delay-and-summed (DAS) beams into the approximate post-processed images found on clinical-grade scanners. Training MimickNet only requires post-processed image samples from a scanner of interest without the need for explicit pairing to raw DAS data. This flexibility allows it to hypothetically approximate any manufacturer's post-processing without access to the pre-processed data. MimickNet generates images with an average similarity index measurement (SSIM) of 0.930$\pm$0.0892 on a 300 cineloop test set, and it generalizes to cardiac cineloops outside of our train-test distribution achieving an SSIM of 0.967$\pm$0.002. We also explore the theoretical SSIM achievable by evaluating MimickNet performance when trained under gray-box constraints (i.e., when both pre-processed and post-processed images are available). To our knowledge, this is the first work to establish deep learning models that closely approximate current clinical-grade ultrasound post-processing under realistic black-box constraints where before and after post-processing data is unavailable. MimickNet serves as a clinical post-processing baseline for future works in ultrasound image formation to compare against. To this end, we have made the MimickNet software open source.
\end{abstract}

%
\IEEEpeerreviewmaketitle
\global\csname @topnum\endcsname 0
\global\csname @botnum\endcsname 0

\section{Introduction and Background}
\begin{figure}
    \includegraphics[width={\linewidth}]{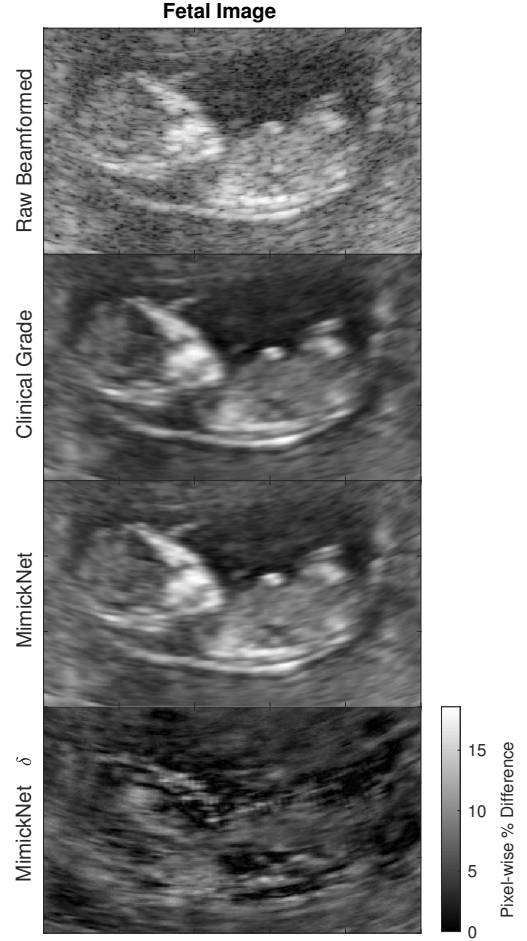}
    \caption{Fetal image comparing clinical-grade post-processed images (ground truth) and MimickNet post-processing. In the last row, the difference between clinical-grade and MimickNet post-processing is scaled to maximize dynamic range. The SSIM of the MimickNet image to clinical grade image is 0.972 and the PSNR is 26.78.}\label{fig:main}
\end{figure}

\IEEEPARstart{I}{n} the typical clinical B-mode ultrasound imaging paradigm, a transducer probe will transmit acoustic energy into tissue, and the back-scatter energy is reconstructed via beamforming techniques into a human eye-friendly image. This image attempts to faithfully map tissue's acoustic impedance, which is a property of its bulk modulus and density. Unfortunately, there are many sources of image degradation such as electronic noise, speckle from sub-resolution scatterers, reverberation, and de-focusing caused by heterogeneity in tissue sound speed \cite{Pinton2011-cs}. In the literature, these sources of image degradation can be suppressed through better focusing \cite{nsight,Bottenus2018-nj}, spatial compounding \cite{Trahey1987-qh}, harmonic imaging \cite{Anvari2015-ei}, and coherence imaging techniques \cite{Long2018-yv,Morgan2019-mh}.

In addition to beamforming, image post-processing is a significant contributor to image quality improvement. Reader studies have shown that medical providers largely prefer post-processed images over raw beamformed imagery \cite{Ahman2009-en,Long2018-yv}. Unfortunately, commercial post-processing algorithms are proprietary, and implementation details are typically kept as a black-box to the end-user. Thus, researchers that develop image improvement techniques on highly configurable research systems, such as Verasonics and Cephasonics scanners, face challenges in presenting their images alongside current clinical system scanner baselines. The current status quo for researchers working on novel image forming techniques is to compare against raw beamformed data which is not typically viewed by medical providers. To have a pixel-wise comparison with clinical-grade standards, researchers would either need access to proprietary post-processing code or access to raw data from difficult-to-configure commercial scanners. We aim to remove these significant barriers by leveraging recent deep learning methods.

Deep learning based post-processing using convolutional neural network (CNN) generators \cite{Xu2014-xt, Mao2016-xg} have become immensely popular in the image restoration problem \cite{Takeda2006-tw,Tomasi1998-tn}. One popular network architecture used is an encoder-decoder network with skip connections commonly referred to as a Unet \cite{Ronneberger2015-kw}. In the image restoration problem, the encoder portion of Unet takes a noisy image as input and creates feature map stacks which are subsequently down-sampled through max pool operations. The decoder portion up-samples features and attempts to reconstruct an image of the same size as the input. Usage of skip connections in Unet has been shown to better maintain high-frequency information in the original image than without \cite{Yamanaka2017-mk}. Other encoder-decoder Unet flavours exist which exploit residual learning \cite{He2015-fc,Zhang2017-qz}, wavelet transforms \cite{Liu2018-rm}, and dense blocks \cite{Huang2016-wk, Jegou2016-nc}. Encoder and decoder network parameters can be optimized typically with a gradient descent based method which minimizes a distance function between the reconstructed and ground truth image \cite{Vincent2010-ea}. Different distance functions such as mean squared error (MSE), mean absolute error (MAE), and structural similarity index measurement (SSIM) have been used in practice \cite{Zhao2015-cc, Snell2015-pz}.

Adversarial objective functions are a unique class of distance functions that have shown success in the related field of image generation \cite{Brock2018-wz}. The adversarial objective optimizes two networks simultaneously. Given training batch sizes of $m$ with individual examples $i$, $G$ is a network that generates images from noise $z^{(i)}$, and another network, $D$, discriminates between real images $x^{(i)}$ and fake generated images $G(z^{(i)})$. $D$ and $G$ play a min-max game since they have competing objective functions shown in Eq. \ref{eq:1} and Eq. \ref{eq:2} where $\theta_{g}$ are parameters of $G$ and $\theta_{d}$ are parameters of $D$. If this min-max game converges, $G$ ultimately learns to generate realistic fake images that are indistinguishable from the perspective of $D$.
\begin{align}
  &\argmin_{\theta_{g}}\frac{1}{m} \sum^{m}_{i=1} \log(1 - D_{\theta_{d}}(G_{\theta_{g}}(z^{(i)}))),\label{eq:1} \\
  &\argmax_{\theta_{d}}\frac{1}{m} \sum^{m}_{i=1} \log D_{\theta_{d}}(x^{(i)}) +\log(1 - D_{\theta_{d}}(G_{\theta_{g}}(z^{(i)}))). \label{eq:2}
\end{align}
In the literature, these networks are referred to as generative adversarial networks (GANs) \cite{Goodfellow2014-ti, Radford2015-qw}. Conditional GANs (cGANs) have seen success in image restoration as well as style transfer. With cGANs, a structured input, such as an image segmentation or corrupted image, is given instead of random noise \cite{Isola2016-wp}.

In the field of ultrasound, deep learning techniques using cGANs and CNNs have recently been applied to B-mode imaging. They have shown promising results for reducing speckle noise, enhancing image contrast, and increasing other image quality metrics \cite{Abdel-Nasser2017-fc, Dietrichson2018-ea, Perdios2018-qe}. However, training GANs or CNNs for image enhancement require ground truths for comparison. These are typically before and after image enhancement pairs that are registered with one another. Unfortunately, this is a luxury not often available in most research environments requiring clinical-grade ground truths.

An extension of GANs known as cycle-consistent GANs (CycleGAN) has been proposed by \cite{Zhu2017-cl} to get around the requirement of paired images. CycleGANs are shown to excel at the problem of style transfer where images are mapped from one domain to another without the use of explicitly paired images. CycleGANs consist of two key components: forward-reverse domain generators, $G_{a}$ and $G_{b}$, and forward-reverse domain discriminators, $D_{a}$ and $D_{b}$. The generators translate images from one domain to another, and the discriminators distinguish between real and fake generated images in each domain. We show the objective functions for one direction of the cycle in Eq. \ref{eq:3} and Eq. \ref{eq:4} where  $a^{(i)}$ is an image from domain $A$, and $b^{(i)}$ is an image from domain $B$. In Eq. \ref{eq:3} and Eq. \ref{eq:4}, the variables $\theta_{g_{a}}$ and $\theta_{d_{b}}$ are the parameters for the domain $A$ forward generator and domain $B$ discriminator. In Eq. \ref{eq:3}, $f$ can represent any distance metric to compare two images.
\begin{align}
  &\argmin_{\theta_{g_{a}}} \; f(G_{a}(G_{b}(a^{(i)})), a^{(i)}), \label{eq:3} \\
  &\argmax_{\theta_{d_{b}}} \frac{1}{m} \sum^{m}_{i=1} \log D_{b}(b^{(i)}) + \log(1 - D_{b}(G_{b}(a^{(i)}))). \label{eq:4}
\end{align}

In this work, we investigate if it is possible to approximate post-processing algorithms found on clinical-grade scanners given raw conventional beamformed data as input to Unet generators. We first show what is theoretically feasible when before and after image pairs are provided and refer to this as a gray-box constraint. We view this as the classic image restoration problem where clinical-grade post-processed images are ground truth, and raw data are ``corrupted". Later, we constrain ourselves to the more realistic black-box setting where no before and after image pairs are available. We view this problem from the style transfer lens and train a CycleGAN from scratch to mimic clinical-grade post-processing. We refer to this trained model configuration as MimickNet. Our results suggest that any manufacturers' post-processing can be well approximated using this framework with just data acquired through a clinical scanner's intended use.

\section{Methodology}
We start with 1500 unique ultrasound image cineloops from fetal, phantom, and liver targets across Siemens S2000, SC2000, or Verasonics Vantage scanners using various scan parameters from \cite{Kakkad2015-wu, Deng2017-ko, Long2018-yv, long2018implications}. This study was approved by the Institutional Review Board at the Duke University, and each study subject provided written informed consent prior to enrollment in the study. We split whole cineloops into respective training and testing sets. Each cineloop has multiple image frames of conventional delay and summed (DAS) beamformed data. The datasets combined consist of 39200 frames with a 30691/8509 image frame train-test split. Each image frame runs through a Siemens proprietary compiled post-processing software producing before and after pairs. These pairs are shuffled and randomly cropped to 512x512 images with padded reflection if the dimensions are too small. Constraining the image dimensions enables batch training, which leads to faster and more stable training convergence. During inference time, images can be any size as long as they are divisible by 16 due to required padding in our CNN architecture. Table \ref{table:dataset} contains details about our training data.

\begin{table}
\caption{Dataset Overview}\label{table:dataset}
\centering 
\begin{tabular}{l|r|rrr}
Scanner Type & Targets & Frames & Train Frames & Test Frames \\ \hline \hline 
\textbf{S2000} & 873 & 3085 & 2543 & 542 \\
\textbf{SC2000} & 158 & 12806 & 9754 & 3052 \\
\textbf{Verasonics} & 469 & 23309 & 18394 & 4915 \\ \hline 
\textbf{Total} & \textbf{1500} & \textbf{39200} & \textbf{30691} & \textbf{8509} \\
\end{tabular}
\end{table}

\subsection{Gray-box Performance with Paired Images}
\begin{align}
  MSE(x,y) &= \frac{1}{m}\sum_{i=1}^{m} (x^{(i)} - y^{(i)})^{2}, \label{eq:mse} \\
  MAE(x,y) &= \frac{1}{m}\sum_{i=1}^{m} |x^{(i)} - y^{(i)}|, \label{eq:mae} \\
  PSNR(x, y) &= 20\log_{10}\frac{MAX_{i}}{\sqrt{MSE}}, \label{eq:psnr}\\
  SSIM(X, Y) &= l(X,Y)*c(X,Y)*s(X,Y), \label{eq:ssim} \\\
  l(X, Y) &= \frac{2\mu_{X}\mu_{Y}+c_{1}}{\mu_{X}^{2}+ \mu_{Y}^{2} + c_{1}}, \label{eq:l_ssim} \\
  c(X, Y) &= \frac{2\sigma_{X}\sigma_{Y} + c_{2}}{\sigma_{X}^{2}+ \sigma_{Y}^{2} + c_{2}}, \label{eq:c_ssim} \\
  s(X, Y) &= \frac{\sigma_{XY} + c_{3}}{\sigma_{X}\sigma_{Y} + c_{3}}. \label{eq:s_ssim}
\end{align}

In the gray-box case where before and after paired images are available, our problem can be seen as a classic image restoration problem where our input DAS beamformed data is ``corrupted", and our clinical-grade post-processed image is the ``uncorrupted ground truth". We optimize for the different distance metrics MSE, MAE, and SSIM. As defined in Eq. \ref{eq:mse}, MSE is the summed pixel-wise squared difference between a ground truth pixel $y^{(i)}$ in image $y$ and estimated pixel $x^{(i)}$ in image $x$. These residuals are averaged by all pixels $m$ in the image. MAE is defined in Eq. \ref{eq:mae} as the summed pixel-wise absolute difference. SSIM is defined in Eq. \ref{eq:ssim} and is the multiplicative similarity between two images' luminance $l$, contrast $c$, and structure $s$ (Eq. \ref{eq:l_ssim}-\ref{eq:s_ssim}). SSIM constants we use are based on \cite{Bovik2004-uc}. $X$ and $Y$ define 11$\times$11 kernels on two images we wish to calculate the similarity of. These kernels slide across the two images, and the output values are averaged to get the SSIM between two images. Variables $\mu_{X}$, $\sigma_{X}^2$ and $\mu_{Y}$, $\sigma_{Y}^2$ are the mean and variance of each kernel patch, respectively. Variables $c_{1}$, $c_{2}$, and $c_{3}$ are the constants $(k_{1}L)^{2}$,  $(k_{2}L)^{2}$, and $c_{2}/2$ respectively. $L$ is the dynamic range of the two images, $k_{1}$ is 0.01, and $k_{2}$ is 0.03.

We calculate SSIM and peak signal to noise ratio (PSNR) by running our trained model on the full test set with images at their original non-padded size. PSNR is defined by Eq. \ref{eq:psnr} where $MAX_{i}$ is the maximum possible intensity of the image.

\subsection{Black-box Performance with Unpaired Images}
To simulate the more realistic black-box case where paired before and after images are unavailable, we take whole cineloops from the training set used in the gray-box case and split them into two groups. For the first group, we only use the raw beamformed data, and for the second group, we only use the clinical-grade post-processed data. We then train a CycleGAN using different distance metrics MSE, MAE, and SSIM for our generators' cycle-consistency loss (Eq. \ref{eq:3}). Like in the gray-box case, MSE, MAE, PSNR, and SSIM metrics were calculated by running our trained model on the full test set to their original non-padded size. Since we have access to the underlying proprietary clinical post-processing, we can compare against objective ground truths solely for final evaluation.

\subsection{Generator and Discriminator Structure}
\begin{figure}[h]
    \includegraphics[width={\linewidth}]{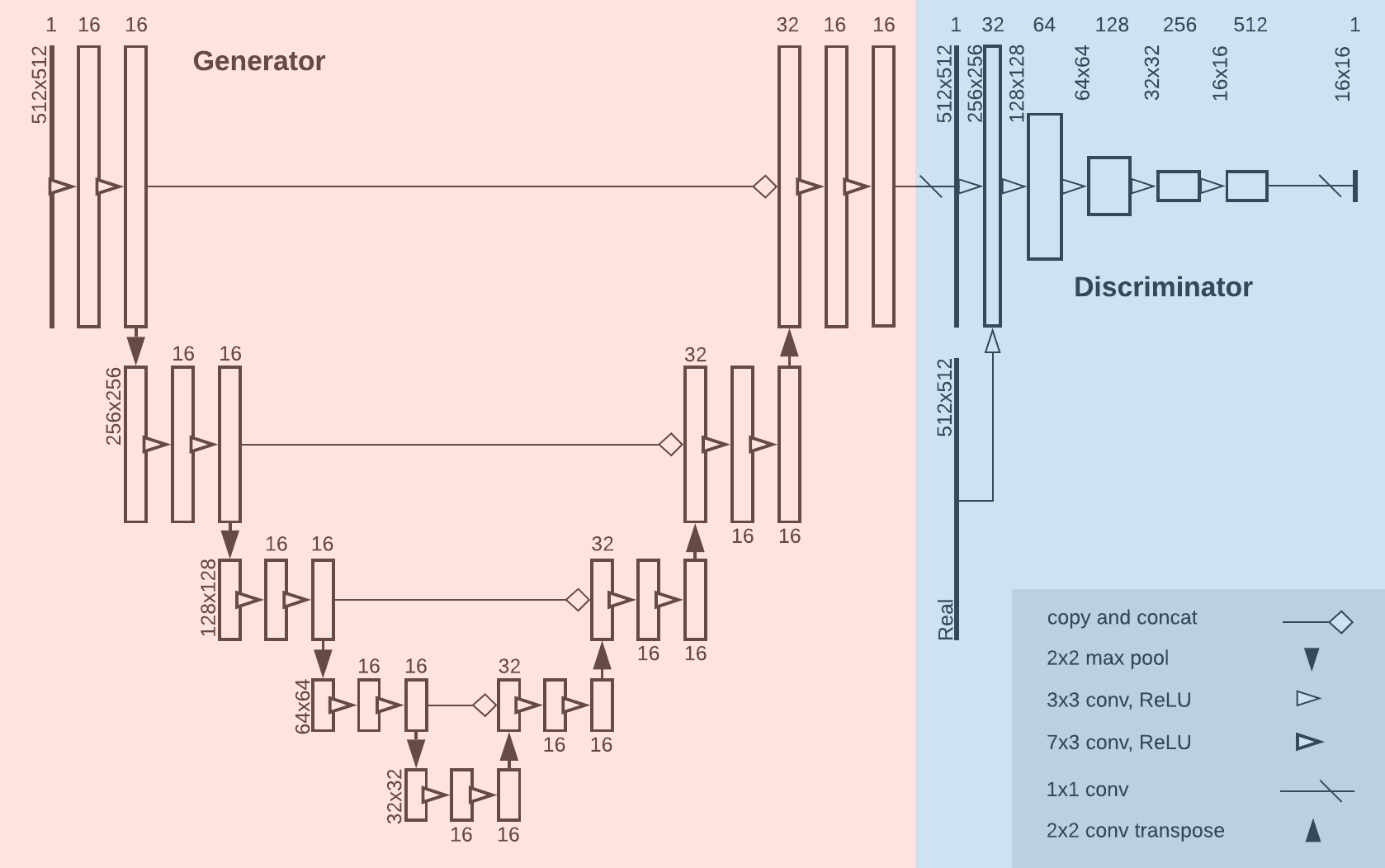}
    \caption{Above is a diagram of the generator and discriminator structure for MimickNet in one translation direction. Note: the reverse translation direction uses an identical mirrored structure. Under gray-box training constraints, only the generator is used. }\label{fig:mimicknet_arch}
\end{figure}

The same overall generator network structure is used in both the gray-box and black-box cases. We use a simple encoder-decoder with skip connections as seen on the left side of Fig. \ref{fig:mimicknet_arch}. We vary filter sizes and the number of filters per layer as hyperparameters to the generator, and we report the total number of weight parameters in each model variation.

The discriminator structure on the right side of Fig. \ref{fig:mimicknet_arch} follows the PatchGAN and LSGAN approach used in \cite{Isola2016-wp, Mao2016-xg} to optimize for least-squares on patches of linearly activated final outputs. The discriminator is only used to facilitate training in the black-box case where no paired images are available, and it is not used in the gray-box case since ground truths are available. Code and models are available at \texttt{https://github.com/ouwen/mimicknet}.
\subsection{Worst Case Performance}
\begin{align}
  cs(X, Y) &= \frac{2\sigma_{XY} + c_{2}}{\sigma_{X}^{2}+ \sigma_{Y}^{2} + c_{2}}. \label{eq:cs_ssim}
\end{align}

We investigate outlier images that perform worst on the SSIM metric by breaking SSIM into its three components: luminance $l$, contrast $c$, and structure $s$. The equations for contrast $c$ and structure $s$ are highly related in examining variance between and within patches. Thus, $c$ and $s$ are simplified into a single contrast-structure $cs$ equation (Eq. \ref{eq:cs_ssim}).

\section{Results}
\subsection{Gray-Box Performance with Paired Images}
In the theoretical gray-box case where before and after paired images are available, we explore different possible Unet encoder-decoder hyperparameters. For each hyperparameter variation, we trained a triplet of models that optimize for SSIM, MSE, and MAE. We note that within each triplet, models using the SSIM minimization objective have the best SSIM and PSNR. We are primarily interested in the best SSIM metric since it was originally formulated to model the human visual system \cite{Bovik2004-uc}. In Table \ref{table:paired_performance}, the best average metrics of each column are in bold. Many of the metrics across model variations are not significantly different, but the SSIM for every model is above 0.967. For subsequent worst-case performance analysis, we used the 52993 parameter model optimized on SSIM loss. This model corresponds to the same generator structure used in Fig. \ref{fig:mimicknet_arch} except with a 3$\times$3 instead of a 7$\times$3 filter.

\begin{table}[h]
\caption{gray-box Performance with Paired Images. Best average metrics are emphasized in bold.}\label{table:paired_performance}
\centering 
\resizebox{\linewidth}{!}{%
    \begin{tabular}{llllll}
    \bf{Loss} & \bf{Params} & \bf{MSE $10^{-3}$} & \bf{MAE $10^{-2}$} & \bf{PSNR} & \bf{SSIM} \\ \hline \hline 
    ssim & 13377 & 2.78$\pm$3.22 & 3.97$\pm$2.40 & 27.4$\pm$3.9 & 0.967$\pm$0.015 \\
    mse & 13377 & 2.40$\pm$2.65 & 3.76$\pm$2.00 & 27.7$\pm$3.4 & 0.947$\pm$0.022 \\
    mae & 13377 & 2.51$\pm$2.86 & 3.83$\pm$2.13 & 27.6$\pm$3.5 & 0.946$\pm$0.018 \\ \hline
    ssim & 29601 & 2.63$\pm$3.10 & 3.91$\pm$2.40 & 27.9$\pm$4.0 & 0.967$\pm$0.015 \\
    mse & 29601 & 2.19$\pm$2.25 & 3.61$\pm$1.81 & 27.9$\pm$3.2 & 0.940$\pm$0.019 \\
    mae & 29601 & 3.46$\pm$3.20 & 4.58$\pm$2.16 & 25.7$\pm$3.0 & 0.895$\pm$0.028 \\ \hline
    ssim & 34849 & 2.49$\pm$2.88 & 3.78$\pm$2.28 & 27.9$\pm$3.9 & 0.975$\pm$0.013 \\ 
    mse & 34849 & 2.27$\pm$2.41 & 3.67$\pm$1.92 & 27.9$\pm$3.3 & 0.950$\pm$0.019 \\  
    mae & 34849 & 2.31$\pm$2.54 & 3.68$\pm$1.96 & 27.9$\pm$3.4 & 0.951$\pm$0.016 \\ \hline
    ssim & 52993 & 2.28$\pm$2.77 & 3.65$\pm$2.24 & 28.5$\pm$4.2 & \textbf{0.979$\pm$0.013} \\ 
    mse & 52993 & 2.19$\pm$2.40 & 3.60$\pm$1.92 & 28.1$\pm$3.4 & 0.956$\pm$0.017 \\
    mae & 52993 & 2.11$\pm$2.35 & 3.52$\pm$1.89 & 28.3$\pm$3.4 & 0.959$\pm$0.015 \\  \hline
    ssim & 77185 & 2.38$\pm$2.91 & 3.70$\pm$2.28 & 28.3$\pm$4.0 & 0.976$\pm$0.015 \\ 
    mse & 77185 & \textbf{2.02$\pm$2.09} & \textbf{3.46$\pm$1.70} & 28.3$\pm$3.2 & 0.946$\pm$0.022 \\ 
    mae & 77185 & 2.14$\pm$2.23 & 3.55$\pm$1.80 & 28.0$\pm$3.2 & 0.947$\pm$0.020 \\ \hline
    ssim & 117697 & 2.22$\pm$2.65 & 3.59$\pm$2.11 & 28.4$\pm$3.9 & 0.977$\pm$0.014 \\ 
    mse & 117697 & 2.72$\pm$2.51 & 4.07$\pm$1.95 & 26.9$\pm$3.1 & 0.931$\pm$0.023 \\
    mae & 117697 & 2.93$\pm$2.93 & 4.18$\pm$2.11 & 26.7$\pm$3.3 & 0.927$\pm$0.022 \\ \hline
    ssim & 330401 & 2.25$\pm$2.79 & 3.61$\pm$2.22 & \textbf{28.6$\pm$4.1} & 0.977$\pm$0.013 \\ 
    mse & 330401 & 2.15$\pm$2.20 & 3.58$\pm$1.83 & 28.1$\pm$3.4 & 0.958$\pm$0.016 \\
    mae & 330401 & 2.23$\pm$2.42 & 3.61$\pm$1.89 & 28.0$\pm$3.4 & 0.958$\pm$0.016 \\ \hline
    ssim & 733025 & 2.63$\pm$3.06 & 3.93$\pm$2.33 & 27.7$\pm$4.0 & 0.967$\pm$0.015 \\ 
    mse & 733025 & 2.40$\pm$2.51 & 3.79$\pm$1.97 & 27.7$\pm$3.4 & 0.945$\pm$0.023 \\
    mae & 733025 & 2.80$\pm$2.83 & 4.09$\pm$2.04 & 26.9$\pm$3.2 & 0.927$\pm$0.022 \\
    \end{tabular}
}
\end{table}

\subsection{Black-box Performance with Unpaired Images}
In the more realistic black-box case where before and after images are not available, we also explore different Unet architecture hyperparameters. We attempted to train from scratch the same 52993 parameter generator network architecture selected from Table \ref{table:paired_performance}, but we were unsuccessful in guiding convergence without increasing the number of generator parameters to 117697. This increase was accomplished by changing every filter size from 3$\times$3 to 7$\times$3, and metrics can be seen in Table \ref{table:unpaired_performance}. For the large 7.76M parameter generator network, performance differences between triplets of the objective functions are not significant. The row labeled ``ver", is a model trained only on Verasonics Vantage data with MAE optimization.

\begin{figure}
    \includegraphics[width={\linewidth}]{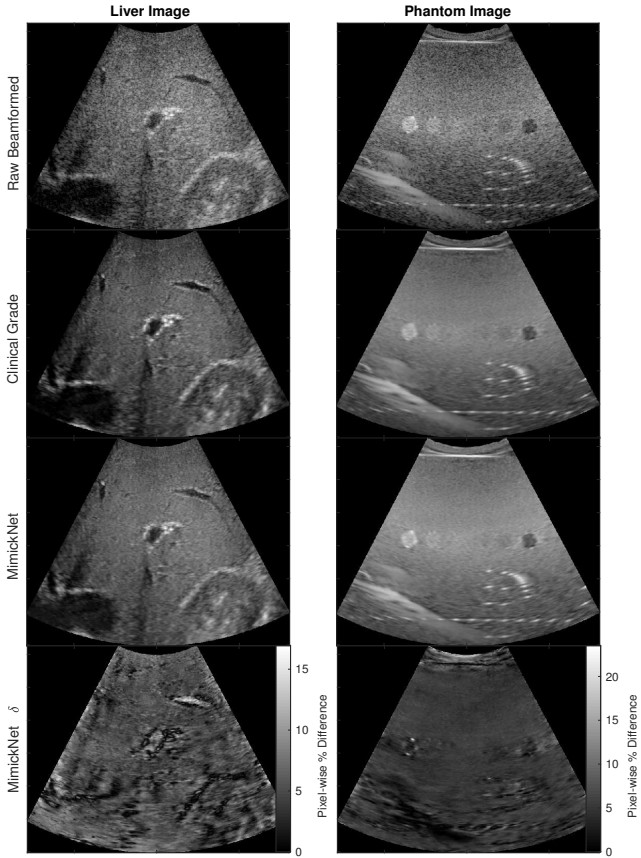}
    \caption{Liver (left) and phantom (right) images. The difference between clinical-grade and MimickNet outputs are scaled to maximize dynamic range. The SSIM and PSNR between MimickNet and clinical-grade images for the liver target is 0.9472 and 26.91, respectively. The SSIM and PSNR between MimickNet and clinical-grade images for the phantom target is 0.9802 and 27.20, respectively.}\label{fig:2}
\end{figure}

We select the 117697 parameter network optimizing MSE for subsequent analysis since it achieves the highest SSIM with fewest parameters. We refer to this configuration, shown in Fig. \ref{fig:mimicknet_arch}, as MimickNet. In Fig. \ref{fig:main} and Fig. \ref{fig:2}, fetal, liver, and phantom images are shown. Without the scaled differences in the last row, it is much more difficult to discern localized differences between MimickNet images and clinical-grade post-processed images.

\begin{table}[h]
\caption{Black-box Performance with Unpaired Images, ``ver", is a model trained only on Verasonics Vantage data with the MAE optimization.}\label{table:unpaired_performance}
\centering
\resizebox{\linewidth}{!}{%
    \begin{tabular}{llllll}
    \bf{Loss} & \bf{Params} & \bf{MSE $10^{-3}$} & \bf{MAE $10^{-2}$} & \bf{PSNR} & \bf{SSIM} \\ \hline \hline 
    ssim & 117697 & 7.26$\pm$10.5 & 6.54$\pm$4.38 & 23.9$\pm$4.41 & 0.883$\pm$0.091 \\
    mse & 117697 & 6.83$\pm$11.1 & 6.31$\pm$4.39 & 24.7$\pm$4.95 & \textbf{0.930$\pm$0.089} \\
    mae & 117697 & 6.79$\pm$9.89 & 6.30$\pm$4.27 & 24.4$\pm$4.67 & 0.900$\pm$0.085 \\  \hline
    ssim & 7.76M & \textbf{4.45$\pm$5.71} & \textbf{5.14$\pm$3.12} & \textbf{25.6$\pm$4.08} & 0.918$\pm$0.078 \\
    mse & 7.76M & 6.23$\pm$6.30 & 6.14$\pm$3.24 & 23.6$\pm$3.57 & 0.897$\pm$0.052 \\ 
    mae & 7.76M & 6.20$\pm$9.10 & 6.02$\pm$4.21 & 25.1$\pm$5.05 & 0.918$\pm$0.084 \\ \hline
    ver  & 7.76M & 6.13$\pm$8.95 & 5.99$\pm$4.19 & 25.2$\pm$5.08 & 0.916$\pm$0.083 \\
    \end{tabular}
}
\end{table}

\subsection{Runtime Performance}
In Table \ref{table:performance}, the runtime was examined for the best SSIM performing model in the gray-box paired image and black-box unpaired image training cases. Frames per second (FPS) measurements were calculated for an NVIDIA P100. Floating-point operations per second (FLOPS) are provided as a hardware independent measurement since runtime generally scales linearly with the number of FLOPS used by the model. As a reference point, we include metrics from MobileNetV2 \cite{8578572}, a lightweight image classifier designed explicitly for use on mobile phones. MimickNet uses 2000x fewer FLOPS compared to MobileNetV2. Note that FPS measurements for MobileNetV2 were performed on a Google Pixel 1 phone from  \cite{8578572} and not an NVIDIA P100.

\begin{table}[H]
\centering 
\caption{Runtime Performance on Nvidia P100 and *Pixel 1 Phone under gray-box and black-box training constraints}\label{table:performance}
\begin{tabular}{lllll}
Model     & Input Size   & Params & MFLOPS & FPS (Hz) \\ \hline \hline 
Gray-box      & 512x512 & 52993  & 0.105  & 142 \\
Black-box (MimickNet)  & 512x512 & 117697 & 0.235  & 92 \\ \hline \hline 
MobileNetV2 & 224x224 & 4.3M   &  569  & 5* \\
\end{tabular}
\end{table}

\subsection{Worst Case Performance}
\begin{figure}
    \includegraphics[width=0.8\linewidth]{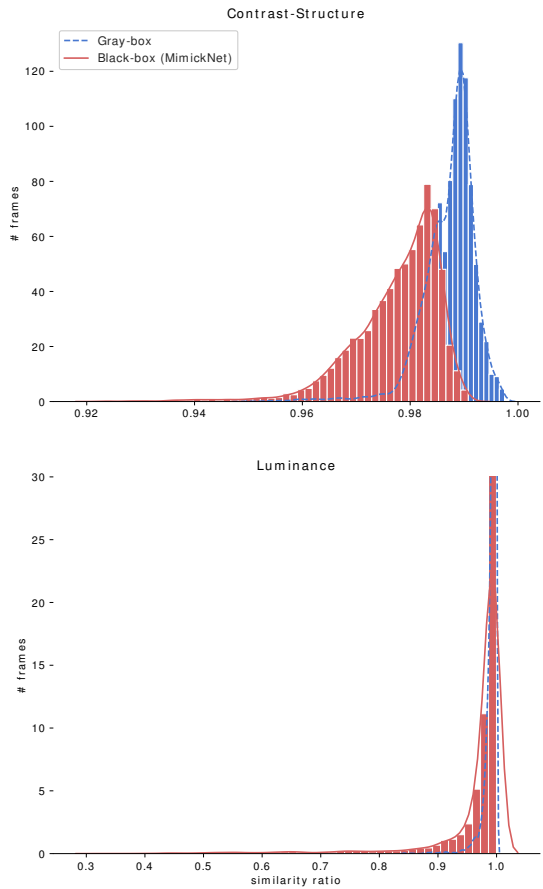}
    \caption{The distribution of contrast-structure (top) and luminance (bottom) of all image frames in our test dataset produced under gray-box and black-box constraints. The $cs$ is $0.987 \pm 0.005$ and $l$ is $0.993 \pm 0.0103$ under gray-box constraints. The $cs$ is $0.978 \pm 0.008$ and $l$ is $0.967 \pm 0.073$ under black-box constraints.  }\label{fig:dist}
\end{figure}

We investigate the distribution of SSIM across our entire test dataset. We break the the SSIM into its luminance $l$ and contrast-structure $cs$ components following Eq. \ref{eq:l_ssim} and Eq. \ref{eq:cs_ssim}. In Fig. \ref{fig:dist}, these components' histogram and kernel density estimate are plotted for the gray-box paired image and the black-box unpaired image training cases. The min-max $cs$ range for the gray-box case is tightly between 0.950 and 0.998, and the black-box case overlaps this region with a min-max $cs$ range between 0.922 and 0.990. The min-max $l$ range of the gray-box case falls between 0.842 and 1.000, but the black-box case has a large min-max range of 0.318 and 1.000.

\begin{figure*}
    \centering
    \includegraphics[width=0.75\textwidth]{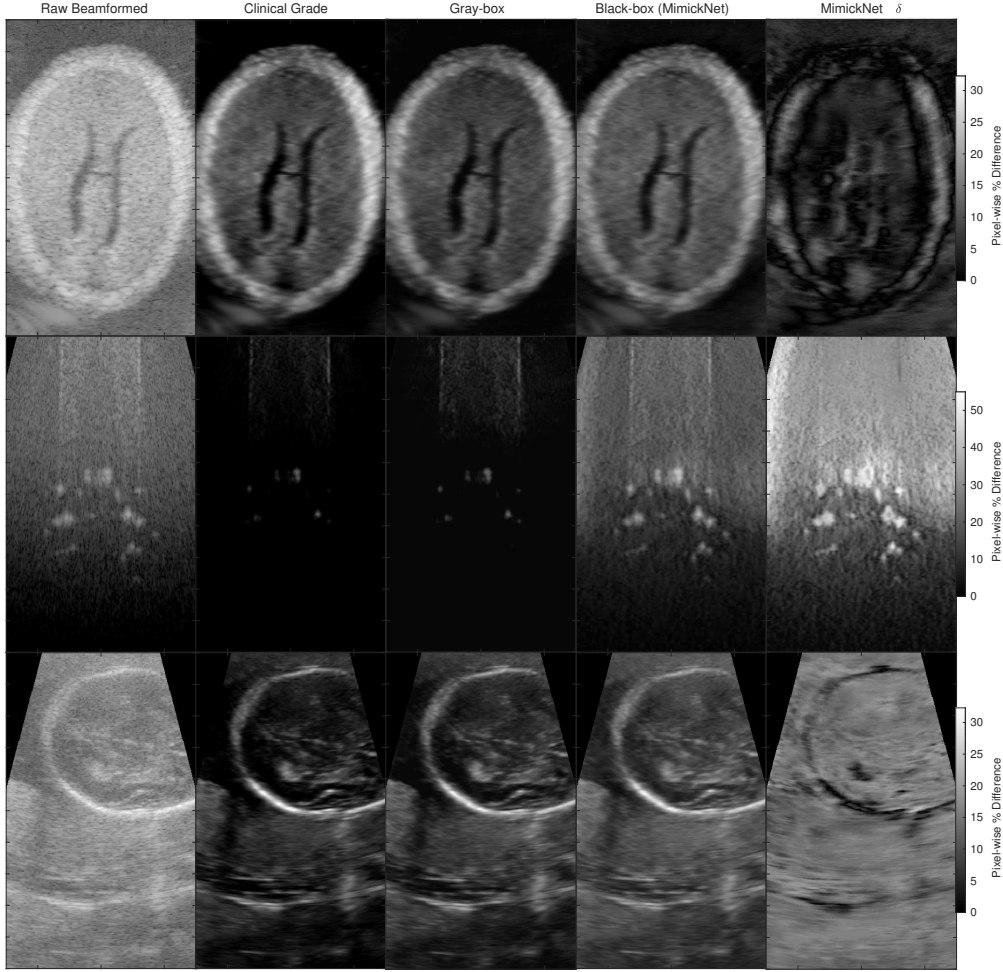}
    \caption{The worst case scenario images for two fetal brain images (top, bottom), and a phantom (middle). The SSIM of the black-box case (MimickNet) to the ground truth images from top to bottom is 0.665 ($cs$ = 0.962, $l$=0.681), 0.414 ($cs$ = 0.947, $l$ = 0.419), and 0.603 ($cs$ = 0.964, $l$ = 0.612). The SSIM of the gray-box case to the ground truth images from top to bottom is 0.873 ($cs$=0.984, $l$=0.883), 0.967 ($cs$=0.996, $l$=0.971), and 0.901 ($cs$=0.988, $l$=0.911). Here $l$ is the luminance and $cs$ is the contrast-structure components of SSIM.}\label{fig:worst}
\end{figure*}

We also closely investigated outlier images that perform poorly on the SSIM metric by looking at the worst images. Fig. \ref{fig:worst} contains three representative images. We included gray-box image results to showcase better the performance gap between what is possible when paired images are available versus when they are not. All three images produced with black-box constraints have high contrast-structure $cs$, but variable luminance $l$.

\subsection{Out of Dataset Distribution Performance}
\begin{figure}
    \includegraphics[width=\linewidth]{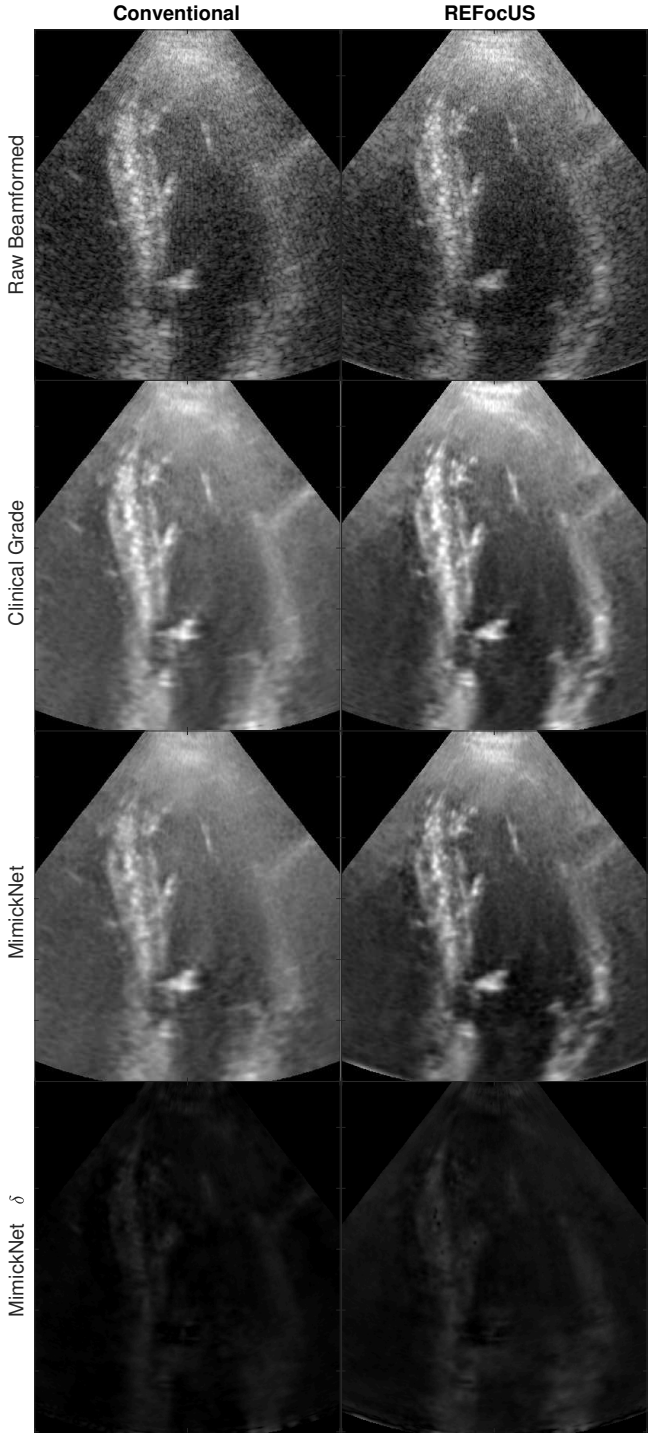}
    \caption{MimickNet applied to out of distribution cardiac data on conventional dynamic receive images and REFocUS ultrasound beamformed images. MimickNet is only trained on fetal, liver, and phantom data. SSIM between clinical-grade post-processing and MimickNet for conventional beamformed images was 0.967$\pm$0.002. SSIM between clinical-grade post-processing and MimickNet for REFocUS beamformed images was 0.950$\pm$0.0157. Note that the last row is at the same scale as the cardiac images above.}\label{fig:cardiac}
\end{figure}

To assess the generalizability of MimickNet post-processing, we applied it to cardiac cineloop data. These data are outside of our train-test dataset distribution which only included phantom, fetal, and liver imaging targets. We also applied MimickNet post-processing to a recent novel beamforming method known as REFocUS \cite{Bottenus2018-nj}. REFocUS allows for transmit-receive focusing everywhere under linear system assumptions resulting in better image resolution and contrast-to-noise ratio. In Fig. \ref{fig:cardiac}, we see that MimickNet post-processed images closely match clinical-grade post-processing for conventional dynamic receive beamforming with an SSIM of 0.967$\pm$0.002. Similar to clinical-grade post-processing, we see that contrast improvements in the heart chamber and resolution improvements along the heart septum due to REFocUS are preserved after MimickNet post-processing, achieving an SSIM of 0.950$\pm$0.0157.

\section{Discussion}
MimickNet can closely approximate clinical-grade post-processing with an SSIM of 0.930$\pm$0.089 such that even upon close inspection, few differences are observed. This performance was achieved without knowledge of the pre-processed pair. We do observe a performance gap compared to the gray-box setting, which achieves an SSIM of 0.979$\pm$0.013. However, emulating the gray-box setting would require researchers to tamper with scanner systems to siphon off pre-processed data, so we explore ways to eliminate this gap.

The performance gap is primarily attributed to differences image luminance from outlier frames seen in Fig. \ref{fig:dist}. Although images generated under black-box constraints present a large min-max $l$ range of 0.318 to 1.000, we note that the mean and standard deviation is $0.967 \pm 0.073$. Therefore, the majority of images do have well-approximated luminance, despite the sizeable min-max range. For the two fetal brain images in Fig. \ref{fig:worst}, we qualitatively see that much of the contrast and structure are preserved while luminance is not. This matches the quantitative contrast-structure and luminance SSIM components for the top fetal image ($cs$ 0.962, $l$ 0.681) and bottom fetal image ($cs$ 0.964, $l$ 0.612).

We found it interesting that clinical-grade post-processing would remove such bright reflectors seen in the raw beamformed phantom image (Fig. \ref{fig:worst}, 2nd row). This level of artifact removal likely requires window clipping. When we clip the lower dynamic range of raw beamformed data from -120dB to -80dB, we see the bright scatterers in raw beamformed images dim and practically match clinical-grade post-processing without any additional changes. Conceptually, clipping values to -80dB is a reasonable choice since it is close to the noise floor of most ultrasound transducers. In the CycleGAN training paradigm, it can be challenging to learn these clipping cutoffs due to the cycle-consistency loss (defined in Eq. \ref{eq:3}). The backward generator would be penalized by any information destroyed through clipping learned in the forward generator. Since the cycle-consistency loss does not exist in optimization under the gray-box setting, the model under gray-box settings can learn the clipping better than under black-box settings. Fortunately, luminance can be modified to a large extent in real-time by changing the imaging window or gain by ultrasound end-users.

One challenge we found was that training MimickNet was quite unstable for small generator networks. This instability is likely due to the nature of adversarial objectives in GANs which other works explore \cite{karras2018progressive, Mescheder2018-ch}. The overall stability of the adversarial objective function appears to be a more important factor in achieving a higher SSIM rather than the specific generator distance metric used such as MAE or SSIM as seen by Table \ref{table:unpaired_performance}. Training GANs is a delicate balancing act between discriminator and generator. If the discriminator overpowers the generator during training, then the generator is unable to outpace the discriminator. A quick solution is to increase the capacity of the generator by adding more parameters, or by decreasing the capacity of the discriminator by taking away parameters until convergence occurs. Future works will explore how to better increase training stability and address any remaining performance gap between the gray-box and black-box constraint settings through different deep learning model architectures, objective functions, or training processes.

As is, MimickNet shows promise for production use. It runs in real-time at 92 FPS on an NVIDIA P100 and uses 2000x fewer FLOPS than models such as MobileNetV2, which was designed for less capable hardware such as mobile phone CPUs. This runtime is relevant since more ultrasound systems are being developed for mobile phone viewing \cite{Hewener2015-jo}. Additionally, the last row of Table \ref{table:unpaired_performance} named ``ver", is a 7.76M parameter model trained only on Verasonics Vantage data with MAE distance metric optimization while achieving similar metrics to training on the full dataset. These results hint at the possibility of achieving similar SSIM with fewer data. Future work will assess the performance of MimickNet on mobile phones and other data or compute constrained settings.

This work's main contribution is in decreasing the barrier of clinical translation for future research. Medical images previously only understood by research domain experts can be translated to clinical-grade images widely familiar to medical providers. Future work will aim to implement a flexible end-to-end software package to train a mimic provided data from two arbitrary scanner systems. Work will also examine how much data is required to create a high-performance mimic.

\section{Conclusion}
MimickNet closely approximates current clinical post-processing in the realistic black-box setting where before and after post-processing image pairs are unavailable. We present it as an image matching tool to provide fair comparisons of novel beamforming and image formation techniques to a clinical baseline mimic. It runs in real-time, works for out-of-distribution cardiac data, and thus shows promise for practical production use. We demonstrated its application in comparing different beamforming methods with clinical-grade post-processing and showed that resolution improvements are carried over into the final post-processed image. Our results with ultrasound data suggest it should also be possible to approximate medical image post-processing in other modalities such as CT and MR.


%

\section*{Acknowledgment}
This work was supported by the National Institute of Biomedical Imaging and Bioengineering under Grant R01-EB026574, and National Institute of Health under Grant 5T32GM007171-44. The authors would like to thank Siemens Medical Inc. USA for in kind technical support.
\ifCLASSOPTIONcaptionsoff
  \newpage
\fi



\bibliographystyle{IEEEtran}
%
\bibliography{refs}

\begin{thebibliography}{10}
\providecommand{\url}[1]{#1}
\csname url@samestyle\endcsname
\providecommand{\newblock}{\relax}
\providecommand{\bibinfo}[2]{#2}
\providecommand{\BIBentrySTDinterwordspacing}{\spaceskip=0pt\relax}
\providecommand{\BIBentryALTinterwordstretchfactor}{4}
\providecommand{\BIBentryALTinterwordspacing}{\spaceskip=\fontdimen2\font plus
\BIBentryALTinterwordstretchfactor\fontdimen3\font minus
  \fontdimen4\font\relax}
\providecommand{\BIBforeignlanguage}[2]{{%
\expandafter\ifx\csname l@#1\endcsname\relax
\typeout{** WARNING: IEEEtran.bst: No hyphenation pattern has been}%
\typeout{** loaded for the language `#1'. Using the pattern for}%
\typeout{** the default language instead.}%
\else
\language=\csname l@#1\endcsname
\fi
#2}}
\providecommand{\BIBdecl}{\relax}
\BIBdecl

\bibitem{Pinton2011-cs}
G.~F. Pinton, G.~E. Trahey, and J.~J. Dahl, ``\BIBforeignlanguage{en}{Sources
  of image degradation in fundamental and harmonic ultrasound imaging using
  nonlinear, full-wave simulations},'' \emph{\BIBforeignlanguage{en}{IEEE
  Trans. Ultrason. Ferroelectr. Freq. Control}}, vol.~58, no.~4, pp. 754--765,
  Apr. 2011.

\bibitem{nsight}
\BIBentryALTinterwordspacing
K.~Thiele, J.~Jago, R.~Entrekin, and R.~Peterson, ``Exploring nsight imaging, a
  totally new architecture for premium ultrasound,'' Philips, Tech. Rep. 4522
  962 95791, June 2013. [Online]. Available:
  \url{https://www.usa.philips.com/healthcare/resources/feature-detail/nsight}
\BIBentrySTDinterwordspacing

\bibitem{Bottenus2018-nj}
N.~Bottenus, ``{REFoCUS}: Ultrasound focusing for the software beamforming
  age,'' in \emph{2018 {IEEE} International Ultrasonics Symposium ({IUS})},
  Oct. 2018, pp. 1--4.

\bibitem{Trahey1987-qh}
G.~E. Trahey, J.~W. Allison, S.~W. Smith, and O.~T. von Ramm, ``Speckle
  reduction achievable by spatial compounding and frequency compounding:
  Experimental results and implications for target detectability,'' in
  \emph{Pattern Recognition and Acoustical Imaging}, vol. 0768.\hskip 1em plus
  0.5em minus 0.4em\relax International Society for Optics and Photonics, Sep.
  1987, pp. 185--192.

\bibitem{Anvari2015-ei}
A.~Anvari, F.~Forsberg, and A.~E. Samir, ``\BIBforeignlanguage{en}{A primer on
  the physical principles of tissue harmonic imaging},''
  \emph{\BIBforeignlanguage{en}{Radiographics}}, vol.~35, no.~7, pp.
  1955--1964, Nov. 2015.

\bibitem{Long2018-yv}
W.~Long, D.~Hyun, K.~R. Choudhury, D.~Bradway, P.~McNally, B.~Boyd,
  S.~Ellestad, and G.~E. Trahey, ``\BIBforeignlanguage{en}{Clinical utility of
  fetal {Short-Lag} spatial coherence imaging},''
  \emph{\BIBforeignlanguage{en}{Ultrasound Med. Biol.}}, vol.~44, no.~4, pp.
  794--806, Apr. 2018.

\bibitem{Morgan2019-mh}
M.~R. Morgan, D.~Hyun, and G.~E. Trahey, ``\BIBforeignlanguage{en}{Short-lag
  spatial coherence imaging in 1.5-d and 1.75-d arrays: Elevation performance
  and array design considerations},'' \emph{\BIBforeignlanguage{en}{IEEE Trans.
  Ultrason. Ferroelectr. Freq. Control}}, Mar. 2019.

\bibitem{Ahman2009-en}
H.~Ahman, L.~Thompson, A.~Swarbrick, and J.~Woodward, ``Understanding the
  advanced signal processing technique of {Real-Time} adaptive filters,''
  \emph{J. Diagn. Med. Sonogr.}, vol.~25, no.~3, pp. 145--160, May 2009.

\bibitem{Xu2014-xt}
L.~Xu, J.~S.~J. Ren, C.~Liu, and J.~Jia, ``Deep convolutional neural network
  for image deconvolution,'' in \emph{Advances in Neural Information Processing
  Systems 27}, Z.~Ghahramani, M.~Welling, C.~Cortes, N.~D. Lawrence, and K.~Q.
  Weinberger, Eds.\hskip 1em plus 0.5em minus 0.4em\relax Curran Associates,
  Inc., 2014, pp. 1790--1798.

\bibitem{Mao2016-xg}
X.~{Mao}, Q.~{Li}, H.~{Xie}, R.~Y.~K. {Lau}, Z.~{Wang}, and S.~P. {Smolley},
  ``Least squares generative adversarial networks,'' in \emph{2017 IEEE
  International Conference on Computer Vision (ICCV)}, Oct 2017, pp.
  2813--2821.

\bibitem{Takeda2006-tw}
H.~Takeda, S.~Farsiu, and P.~Milanfar, ``Robust kernel regression for
  restoration and reconstruction of images from sparse noisy data,'' in
  \emph{2006 International Conference on Image Processing}, Oct. 2006, pp.
  1257--1260.

\bibitem{Tomasi1998-tn}
C.~Tomasi and R.~Manduchi, ``Bilateral filtering for gray and color images,''
  in \emph{Sixth International Conference on Computer Vision ({IEEE} Cat.
  {No.98CH36271})}, Jan. 1998, pp. 839--846.

\bibitem{Ronneberger2015-kw}
O.~Ronneberger, P.~Fischer, and T.~Brox, ``U-net: Convolutional networks for
  biomedical image segmentation,'' \emph{Med. Image Comput. Comput. Assist.
  Interv.}, 2015.

\bibitem{Yamanaka2017-mk}
J.~Yamanaka, S.~Kuwashima, and T.~Kurita, ``Fast and accurate image super
  resolution by deep {CNN} with skip connection and network in network,'' in
  \emph{Neural Information Processing}.\hskip 1em plus 0.5em minus 0.4em\relax
  Springer International Publishing, 2017, pp. 217--225.

\bibitem{He2015-fc}
K.~He, X.~Zhang, S.~Ren, and J.~Sun, ``Deep residual learning for image
  recognition,'' in \emph{Proceedings of the {IEEE} conference on computer
  vision and pattern recognition}, 2016, pp. 770--778.

\bibitem{Zhang2017-qz}
Z.~Zhang, Q.~Liu, and Y.~Wang, ``Road extraction by deep residual {U-Net},''
  \emph{IEEE Geoscience and Remote Sensing Letters}, vol.~15, no.~5, pp.
  749--753, May 2018.

\bibitem{Liu2018-rm}
P.~Liu, H.~Zhang, K.~Zhang, L.~Lin, and W.~Zuo, ``Multi-level {Wavelet-CNN} for
  image restoration,'' 2018.

\bibitem{Huang2016-wk}
G.~Huang, Z.~Liu, L.~Van Der~Maaten, and K.~Q. Weinberger, ``Densely connected
  convolutional networks,'' in \emph{Proceedings of the {IEEE} conference on
  computer vision and pattern recognition}, 2017, pp. 4700--4708.

\bibitem{Jegou2016-nc}
S.~J{\'e}gou, M.~Drozdzal, D.~Vazquez, A.~Romero, and Y.~Bengio, ``The one
  hundred layers tiramisu: Fully convolutional densenets for semantic
  segmentation,'' in \emph{Proceedings of the {IEEE} Conference on Computer
  Vision and Pattern Recognition Workshops}, 2017, pp. 11--19.

\bibitem{Vincent2010-ea}
P.~Vincent, H.~Larochelle, I.~Lajoie, Y.~Bengio, and P.-A. Manzagol, ``Stacked
  denoising autoencoders: Learning useful representations in a deep network
  with a local denoising criterion,'' \emph{J. Mach. Learn. Res.}, vol.~11, no.
  Dec, pp. 3371--3408, 2010.

\bibitem{Zhao2015-cc}
H.~{Zhao}, O.~{Gallo}, I.~{Frosio}, and J.~{Kautz}, ``Loss functions for image
  restoration with neural networks,'' \emph{IEEE Transactions on Computational
  Imaging}, vol.~3, no.~1, pp. 47--57, March 2017.

\bibitem{Snell2015-pz}
J.~Snell, K.~Ridgeway, R.~Liao, B.~D. Roads, M.~C. Mozer, and R.~S. Zemel,
  ``Learning to generate images with perceptual similarity metrics,'' 2017.

\bibitem{Brock2018-wz}
\BIBentryALTinterwordspacing
A.~Brock, J.~Donahue, and K.~Simonyan, ``Large scale {GAN} training for high
  fidelity natural image synthesis,'' in \emph{International Conference on
  Learning Representations}, 2019. [Online]. Available:
  \url{https://openreview.net/forum?id=B1xsqj09Fm}
\BIBentrySTDinterwordspacing

\bibitem{Goodfellow2014-ti}
I.~Goodfellow, J.~Pouget-Abadie, M.~Mirza, B.~Xu, D.~Warde-Farley, S.~Ozair,
  A.~Courville, and Y.~Bengio, ``Generative adversarial nets,'' in
  \emph{Advances in Neural Information Processing Systems 27}, Z.~Ghahramani,
  M.~Welling, C.~Cortes, N.~D. Lawrence, and K.~Q. Weinberger, Eds.\hskip 1em
  plus 0.5em minus 0.4em\relax Curran Associates, Inc., 2014, pp. 2672--2680.

\bibitem{Radford2015-qw}
A.~Radford, L.~Metz, and S.~Chintala, ``Unsupervised representation learning
  with deep convolutional generative adversarial networks,'' \emph{CoRR}, vol.
  abs/1511.06434, 2016.

\bibitem{Isola2016-wp}
P.~Isola, J.-Y. Zhu, T.~Zhou, and A.~A. Efros, ``Image-to-image translation
  with conditional adversarial networks,'' in \emph{Proceedings of the {IEEE}
  conference on computer vision and pattern recognition}, 2017, pp. 1125--1134.

\bibitem{Abdel-Nasser2017-fc}
M.~Abdel-Nasser and O.~A. Omer, ``Ultrasound image enhancement using a deep
  learning architecture,'' in \emph{Proceedings of the International Conference
  on Advanced Intelligent Systems and Informatics 2016}.\hskip 1em plus 0.5em
  minus 0.4em\relax Springer International Publishing, 2017, pp. 639--649.

\bibitem{Dietrichson2018-ea}
F.~Dietrichson, E.~Smistad, A.~Ostvik, and L.~Lovstakken, ``Ultrasound speckle
  reduction using generative adversial networks,'' in \emph{2018 {IEEE}
  International Ultrasonics Symposium ({IUS})}, Oct. 2018, pp. 1--4.

\bibitem{Perdios2018-qe}
D.~Perdios, M.~Vonlanthen, A.~Besson, F.~Martinez, and J.-P. Thiran, ``Deep
  convolutional neural network for ultrasound image enhancement,'' in
  \emph{2018 {IEEE} International Ultrasonics Symposium ({IUS})}, Oct. 2018,
  pp. 1--4.

\bibitem{Zhu2017-cl}
J.-Y. Zhu, T.~Park, P.~Isola, and A.~A. Efros, ``Unpaired image-to-image
  translation using cycle-consistent adversarial networks,'' in
  \emph{Proceedings of the {IEEE} international conference on computer vision},
  2017, pp. 2223--2232.

\bibitem{Kakkad2015-wu}
V.~Kakkad, J.~Dahl, S.~Ellestad, and G.~Trahey, ``\BIBforeignlanguage{en}{In
  vivo application of short-lag spatial coherence and harmonic spatial
  coherence imaging in fetal ultrasound},''
  \emph{\BIBforeignlanguage{en}{Ultrason. Imaging}}, vol.~37, no.~2, pp.
  101--116, Apr. 2015.

\bibitem{Deng2017-ko}
Y.~Deng, M.~L. Palmeri, N.~C. Rouze, G.~E. Trahey, C.~M. Haystead, and K.~R.
  Nightingale, ``\BIBforeignlanguage{en}{Quantifying image quality improvement
  using elevated acoustic output in {B-Mode} harmonic imaging},''
  \emph{\BIBforeignlanguage{en}{Ultrasound Med. Biol.}}, vol.~43, no.~10, pp.
  2416--2425, Oct. 2017.

\bibitem{long2018implications}
J.~Long, W.~Long, N.~Bottenus, G.~F. Pintonl, and G.~E. Trahey, ``Implications
  of lag-one coherence on real-time adaptive frequency selection,'' in
  \emph{2018 IEEE International Ultrasonics Symposium (IUS)}.\hskip 1em plus
  0.5em minus 0.4em\relax IEEE, 2018, pp. 1--9.

\bibitem{Bovik2004-uc}
A.~C. Bovik, H.~R. Sheikh, and E.~P. Simoncelli, ``Image quality assessment:
  from error visibility to structural similarity,'' \emph{IEEE Trans. Image
  Process.}, vol.~13, no.~4, pp. 600--612, Apr. 2004.

\bibitem{8578572}
M.~{Sandler}, A.~{Howard}, M.~{Zhu}, A.~{Zhmoginov}, and L.~{Chen},
  ``Mobilenetv2: Inverted residuals and linear bottlenecks,'' in \emph{2018
  IEEE/CVF Conference on Computer Vision and Pattern Recognition}, June 2018,
  pp. 4510--4520.

\bibitem{karras2018progressive}
\BIBentryALTinterwordspacing
T.~Karras, T.~Aila, S.~Laine, and J.~Lehtinen, ``Progressive growing of {GAN}s
  for improved quality, stability, and variation,'' in \emph{International
  Conference on Learning Representations}, 2018. [Online]. Available:
  \url{https://openreview.net/forum?id=Hk99zCeAb}
\BIBentrySTDinterwordspacing

\bibitem{Mescheder2018-ch}
\BIBentryALTinterwordspacing
L.~Mescheder, A.~Geiger, and S.~Nowozin, ``Which training methods for {GAN}s do
  actually converge?'' in \emph{Proceedings of the 35th International
  Conference on Machine Learning}, ser. Proceedings of Machine Learning
  Research, J.~Dy and A.~Krause, Eds., vol.~80.\hskip 1em plus 0.5em minus
  0.4em\relax Stockholmsmässan, Stockholm Sweden: PMLR, 10--15 Jul 2018, pp.
  3481--3490. [Online]. Available:
  \url{http://proceedings.mlr.press/v80/mescheder18a.html}
\BIBentrySTDinterwordspacing

\bibitem{Hewener2015-jo}
H.~Hewener and S.~Tretbar, ``Mobile ultrafast ultrasound imaging system based
  on smartphone and tablet devices,'' in \emph{2015 {IEEE} International
  Ultrasonics Symposium ({IUS})}, Oct. 2015, pp. 1--4.

\end{thebibliography}

\vfill

\enlargethispage{-5in}

\end{document}